\documentstyle[12pt]{article}
\begin{document}
%%%%%%%%%%%%%%%%%%%%%%%%%%%%
% \begin{flushright}
%    hep-ph/9606366
% \end{flushright}
%
%   TITLE & AUTHORS
%
%%%%%%%%%%%%%%%%%%%%%%%%%%%%%%
\title{\Large\bf 
What information can we obtain from the yield ratio $\pi^-/\pi^+$ 
in heavy-ion collisions ?}
\author{T.Osada$^1$%\thanks{Permanent address: Department of Physics, 
%Tohoku University, Sendai, Japan.}
%~\thanks{e-mail: osada@azusa.shinshu-u.ac.jp.}~,
~S.Sano$^1$,~M.Biyajima$^1$%\thanks{e-mail: minoru44@jpnyitp.bitnet.}
~and G.Wilk$^2$\\%\thanks{e-mail: wilk@fuw.edu.pl.}~\\
{\small $^1$Department of Physics, Faculty of Science, Shinshu University, 
Matsumoto 390, Japan}\\
{\small $^2$Soltan Institute for Nuclear Studies, Zd-PVIII, Ho\.za 69, 
PL-00-681 Warsaw, Poland}\\}
\date{\today}
\maketitle

\begin{abstract}
The recently reported data on the yield ratio $\pi^-/\pi^+$ in
central rapidity region of heavy-ion collisions are analyzed by  
theoretical formula which accounts for Coulomb interaction between 
central charged fragment (CCF) consisting of nearly stopped nucleons
with effective charge $Z_{\mbox{\scriptsize eff}}$ and charged pions
produced in the same region of the phase space. The Coulomb wave
function method is used instead of the usual Gamow factor in order to
account for the finite production range of pions, $\beta$. For
Gaussian shape of the pion production sources it results in a
quasi-scaling in $\beta$ and $Z_{\mbox{\scriptsize eff}}$ which makes
determination of parameters $\beta$ and $Z_{\mbox{\scriptsize eff}}$
from the existing experimental data difficult. Only sufficiently
accurate data taken in the extreme small $m_{\scriptscriptstyle
T}$-$m_{\pi}$ region, where this quasi-scaling is broken, could be
used for this purpose. \\

\end{abstract}
\vspace*{0.75cm}

The ratios of pionic yields $\pi^-/\pi^+$ in
the central rapidity region in high energy heavy-ion collisions have
been recently reported by E866 \cite{E866AuAu} and by NA44
\cite{NA44PbPb} Collaborations. In both cases a strong excess of
yield of $\pi^-$ over that of $\pi^+$ in the low transverse mass
region has been found. Although its theoretical origin is still point
of debate, the most obvious reason to be checked at first instance is
the effect of the Coulomb final state interaction taking place
between charged pions and central charged fragment (CCF) consisting
of nearly stopped nucleons with  effective charge
$Z_{\mbox{\scriptsize eff}}$ and produced in the central rapidity
region \cite{Shuryak}, cf. Fig. 1. In the approximation neglecting
the possible effect of the finite range of the production of pions as
seen, for example, in Bose-Einstein correlation experiments, the
observed results are given by the ratio of Gamow factors only
\cite{Shuryak}:  
\begin{eqnarray}
N^{\pi^{+}}(\mbox{p}_{\mbox{{\scriptsize{rel}}}})
/N^{\pi^{-}}(\mbox{p}_{\mbox{{\scriptsize{rel}}}}) 
=G(\eta)/G(-\eta),\label{GAMOW-FORMULA}
\end{eqnarray}
where Gamow factor $G(\eta)$ is defined as
$G(\eta)=2\pi\eta~/\big(\exp(2\pi\eta)-1\big)$ and
$\eta=Z_{\mbox{\scriptsize eff}}~m_{\pi}\alpha/ 
%|~{\bf p}_{\mbox{{\scriptsize{\bf rel}}}}~|
\mbox{p}_{\mbox{{\scriptsize{rel}}}}$. Here $\alpha$, $m_{\pi}$ and
${\bf p}_{\mbox{{\scriptsize rel}}}$ are the fine structure
constant, mass of $\pi$ meson and its relative momentum in the
two-body system ($\pi$-CCF), respectively. The energies considered
here are high enough to neglect the influences of charged spectators
produced in the fragmentation regions, cf. Fig. 1.\\  

Such problems have been already considered before but either at much
lower energies \cite{GYULASSY,LI} or with the use of Gamow factor
only \cite{WONG}. In this work we shall analyse high energy data and
use the Coulomb wave function method as developed in \cite{Biyaj95-1}
instead of simple Gamow factors to describe the final state Coulomb
interactions. This allows us to account for the finite size of the
emitting source (or, equivalently, for the finiteness of the pion
production range) characterized by parameter $\beta$. The question we
are interested in is: can one obtain from the observed yields of
$\pi^-/\pi^+$ production mentioned before any valuable information
about {\it both} the effective charge of CCF, $Z_{\mbox{\scriptsize
eff}}$, and the range of interaction parameter $\beta$ ? We derive
theoretical formula for the pion production yield based on the
Coulomb wave function convoluted with some pionic source function 
$\rho$(r) and apply it to the analysis of recent experimental data on
Au+Au collisions at $11.0$ GeV/nucleon \cite{E866AuAu} and on Pb+Pb
Collisions at $158$ GeV/nucleon \cite{NA44PbPb}. The emerging
valley-like structures in the map of the $\chi^2$-values made in
$\beta$-$Z_{\mbox{\scriptsize eff}}$ parameter space is shown to be
connected with the quasi-scaling behaviour of integrals of the square
of Coulomb wave function convoluted with pion production source
function \cite{F12}. This finding shows that the expected
determination of parameters $\beta$ and $Z_{\mbox{\scriptsize eff}}$
from the existing experimental data is difficult if not impossible. \\

Let us consider Coulomb interaction between $\pi^+$ with lab rapidity
$y_{\pi}$ and CCF of some effective mass $M_{\mbox{\scriptsize eff}}$
and with rapidity $Y_{\mbox{\scriptsize CM}}$ (which we assume to be
equal to the cms rapidity of the colliding system). In the rest frame
of CCF the $\pi^+$  momentum  ${\bf p}_{\pi}$ is given by 
\begin{eqnarray}
{\bf p}_{\pi}&\!\!\equiv\!\!&(~\mbox{p}_{\scriptscriptstyle L},
=\big(~m_{\scriptscriptstyle T} \sinh (y_{\pi}-Y_{\mbox{\scriptsize CM}})~
,~{\bf p}_{\scriptscriptstyle T}~\big)\label{PI-MOMENTUM}
\end{eqnarray}
where $m_{\scriptscriptstyle T}=\sqrt{m_{\pi}^2 + {\bf
p}^2_{\scriptscriptstyle T}}$ and  ${\bf p}_{\scriptscriptstyle T}$
is its transverse momentum. In this frame the Schr\"odinger
equation for the relative motion of the two-body system ($\pi^+$-CCF)
is given by
\begin{eqnarray}   
\bigg[~~\frac{\widehat{{\bf p}}^2_r}{2\mu}
+\frac{Z_{\mbox{\scriptsize eff}}~e^2}{r}~~\bigg]\psi_r({\bf r})
=E_{r}\psi_r({\bf r}),
\end{eqnarray}
where ${\bf r}$ is the relative coordinate of $\pi^+$ in the rest
frame of the CCF, $\widehat{{\bf p}}_r \equiv-i\hbar\nabla_r$ and
$\mu$ is the reduced mass of our two-body system:
$\mu=m_{\pi}M_{\mbox{\scriptsize eff}}/(M_{\mbox{\scriptsize
eff}}+m_{\pi})\approx m_{\pi}$ \cite{F1}. One can find that the wave
function $\psi_r$ is given by the following confluent
hypergeometric function \cite{Biyaj95-1,Schiff}:
\begin{eqnarray}
&&\psi_r({\bf p}_{r},{\bf r})=\nonumber \\
&&\Gamma(1+i\eta)e^{-\pi\eta/2}e^{i{\bf p}_r\cdot{\bf r}}
F\big(-i\eta,1,i~(\mbox{p}_{r}r-{\bf p}_{r}\cdot{\bf r})\big)~ \label{CONFHYP}
\end{eqnarray}
where
${\bf p}_{r}=
M_{\mbox{\scriptsize eff}}
~{\bf p}_{\pi}/(M_{\mbox{\scriptsize eff}}+m_{\pi})$.
Convoluting now eq.(\ref{CONFHYP}) with some source function
$\rho({\bf r})$ (i.e., distorting the Coulomb wave function
accordingly in order to account for the finite production range of 
pions), we can now calculate single particle spectra of produced
$\pi^+$'s: 
\begin{eqnarray}
\lefteqn{N^{\pi^{+}}({\bf p}_r;
\eta_+,\beta)=%&\!\!=\!\!&
\int_{}^{}d^3{\bf r}~\rho({\bf r})~\Big|
\psi_r({\bf p}_r,{\bf r})\Big|^2 =}\nonumber \\
&&G(\eta_{+}) \sum_{n=1}^{\infty}\sum_{m=1}^{\infty}
\frac{(-i)^n(i)^m}{n+m+1}
A_n(\eta_{+})A_m^*(\eta_{+})\nonumber \\
&&\times I_R(n,m)~(2\mbox{ p}_r)^{n+m}
\label{2-body}, 
\end{eqnarray}
where 
%\begin{eqnarray*} 
$A_n(\eta_{+})=\Gamma(i\eta_{+}+n)/\big(\Gamma(i\eta_{+})(n!)^2\big)$ 
and 
$\eta_+=+Z_{\mbox{\scriptsize eff}}~\mu\alpha
/\mbox{p}_{\mbox{\scriptsize r}}$. 
%\end{eqnarray*} 
Assuming now that $\rho(r)$ is given by Gaussian distribution:
$\rho(r)=\left( \frac{1}{\sqrt{2 \pi } \beta}\right)^3
\exp\left(\frac{-r^2}{2 \beta^2}\right)$, in which case  
\begin{eqnarray*} 
I_R(n,m)&\!\!=\!\!&4\pi\int_{0}^{\infty} dr r^{n+m+2}\rho(r)\\ 
&\!\!=\!\!&\frac{2}{\sqrt{\pi}}\big(\sqrt{2}\beta~\big)^{n+m}
\Gamma\left(\frac{n+m+3}{2}\right)
\end{eqnarray*} 
(cf. $\Delta_{1C}$ in ref.\cite{Biyaj95-1}), and using the following
decomposition of $Q^2$:
\begin{eqnarray*}
Q^2=\mbox{p}_{\scriptscriptstyle L}^2+{\bf p}^2_{\scriptscriptstyle T}
=\mbox{p}_{\scriptscriptstyle L}^2+(m_{\scriptscriptstyle T}-
                 m_{\pi})^2+2m_{\pi}(m_{\scriptscriptstyle T}-m_{\pi}), 
\end{eqnarray*}
we finally obtain that
\begin{eqnarray}
\lefteqn{N^{\pi^+}(m_t-m_{\pi};\eta_+,\beta)~\Big|
_{\mbox{\scriptsize fixed $y_{\pi}$}}\hspace*{-5.5mm}=G(\eta_{+})
\sum_{n=1}^{\infty}\sum_{m=1}^{\infty}
\frac{(-i)^n(i)^m}{n+m+1}}\nonumber \\
&&\frac{4}{\sqrt{\pi}}
A_n(\eta_{+})A_m^*(\eta_{+})
\Gamma\left(\frac{n+m+3}{2}\right)
(\sqrt{2}\beta~)^{n+m}\nonumber \\ 
&&\times\big(
\mbox{p}_{\scriptscriptstyle L}^2+(m_{\scriptscriptstyle T}-m_{\pi})^2
+2m_{\pi}(m_{\scriptscriptstyle T}-m_{\pi})~\big)
^{(n+m)/2} . \label{WAVE-INTEGRAL}
\end{eqnarray}
For the $\pi^-$ production case the corresponding $N^{\pi^{-}}$ yield
can be obtained by simply changing the sign of the Sommerfeld
parameter in eq.(\ref{WAVE-INTEGRAL}):  $\eta^+ \rightarrow - \eta_+
= \eta_- = -Z_{\mbox{\scriptsize eff}}~\mu\alpha/Q$. Therefore
theoretical formula for the ratio of the production yields we are
looking for is given by: 
\begin{eqnarray}
\pi^-/\pi^+=
\frac{N^{\pi^{-}}(m_{\scriptscriptstyle T}-m_{\pi};\eta_-,\beta)}
{N^{\pi^{+}}(m_{\scriptscriptstyle T}-m_{\pi};\eta_+,\beta)}. \label{FINAL}
\end{eqnarray}

Using now eq.(\ref{FINAL}) we
analyse data on the yield ratios $\pi^+/\pi^-$ observed in Au+Au
collisions \cite{E866AuAu} and on $\pi^-/\pi^+$ ratio observed in
Pb+Pb collisions \cite{NA44PbPb}. For Au+Au collisions data
$Y_{\mbox{\scriptsize CM}}$=1.58 and $y_{\pi}$=2.10 in
eq.(\ref{PI-MOMENTUM}) whereas for data on Pb+Pb collisions
$Y_{\mbox{\scriptsize CM}}$=2.90 and for $y_{\pi}$ we use the
averaged value over rapidities of $\pi$ employed in the respective
data analysis. Applying the minimum $\chi^2$-fitting method when
comparing eq.(\ref{FINAL}) with experimental data one 
discovers (after using a lot of CPU-time) the valley-like
structures in the maps of the   
$\chi^2$-values made in $\beta$-$Z_{\mbox{\scriptsize eff}}$ 
parameter space, cf. Figs. 2a and 2b for E866 and NA44 data,
respectively \cite{F2}.
The values of $\chi^2$ along these valleys are almost constants and equal
to $\chi^2\approx$~35/30 and $\chi^2\approx$~52/53 for E866 and NA44
data, respectively. The traces of the minimum $\chi^2$-values are
almost linear.\\ 

These results strongly suggest the following quasi-scaling behaviour
being present in eq.(\ref{WAVE-INTEGRAL}):  
\begin{eqnarray}
\lefteqn{N^{\pi^+}(m_{\scriptscriptstyle T}-m_{\pi}; 
~\lambda\!\times\!Z_{\mbox{\scriptsize eff }},~\lambda\!\times\!\beta) }
\nonumber \\
&&~\approx~ N^{\pi^+}(m_{\scriptscriptstyle T}-m_{\pi}
;~Z_{\mbox{\scriptsize eff}},~\beta),\label{SCALING} 
\end{eqnarray} 
where $\lambda \!>\!0$. Fig. 3 showing the results of calculations of
eq.(\ref{SCALING}) fully confirms this supposition: both for
$N^{\pi^+}$ and $N^{\pi^-}$ and also for their ratio
$N^{\pi^-}/N^{\pi^+}$. It means, and it is demonstrated in Fig. 4 for
both sets of data analysed here, that present experimental data on
yields $\pi^-/\pi^+$ are not able to determine parameters
$(\beta,Z_{\mbox{\scriptsize eff }})$ uniquely. As one can see we can
explain E866 data by two sets of parameters:
$(\beta,Z_{\mbox{\scriptsize eff}})$=(1.0~fm, 24) and (3.0~fm,72).
Similarly NA44 data can be also described by two sets of parameters:
$(\beta,Z_{\mbox{\scriptsize eff}})$=(2.5~fm, 40) and (5.0~fm,80).
Only at extreme small $m_{\scriptscriptstyle T}-m_{\pi}$
region, where such quasi-scaling in eq.(\ref{SCALING}) seems to be
violated (cf. Fig. 5), such unique determination could be (in
principle) possibly achieved (but only for sufficiently accurate
data).\\ 

Summarizing: we have analyzed recent high energy data for yield
ratios $\pi^-/\pi^+$  measured in the central rapidity regions using
our theoretical formula ({\ref{FINAL}}) calculated by convoluting the
square of Coulomb wave function with Gaussian source function for
pion production. In this way we have accounted for the distortion of
Coulomb wave function caused by the finite size of the production
region. With this formula we find the valley-like structures for the
maps of $\chi^2$-values in the $\beta$-$Z_{\mbox{\scriptsize eff}}$
parameter space which can be attributed to the quasi-scaling
behaviour of eq.(\ref{SCALING}) \cite{F12}. Results obtained by using
only Gamow factor \cite{Shuryak} do not show this property. It means,
as was show in Fig. 4, that it is difficult, if not impossible, to
determine parameters $\beta$ and $Z_{\mbox{\scriptsize eff}}$
uniquely in this  case. They could be determined, if at all, only by
using very accurate data obtained in the extreme small
$m_{\scriptscriptstyle T}-m_{\pi}$ region where this quasi-scaling
property in eq.(\ref{SCALING}) is broken (cf. Fig.5). \\ 

%% ######################################################################

\noindent
{\bf Acknowledgements:}~~~
The authors would like to thank A.~Sakaguchi, T.~Sugitate, N.~Xu and 
H.~Hamagaki for providing their experimental data. Numerical
computations are partially done on the computer at Bubble Chamber
Physics Laboratory (Tohoku University). One of the authors (T.O.)
would like to thank many people who supported him at 
Department of Physics of Shinshu University.
This work is partially supported by Japanese
Grant-in-Aid for Science  Research from the Ministry of Education,
Science and Culture (\#.~06640383).  

%\newpage
%\vspace*{0.25cm}

%\newpage
\begin{center}
{\bf Figure Captions.}
\end{center}

\begin{description}
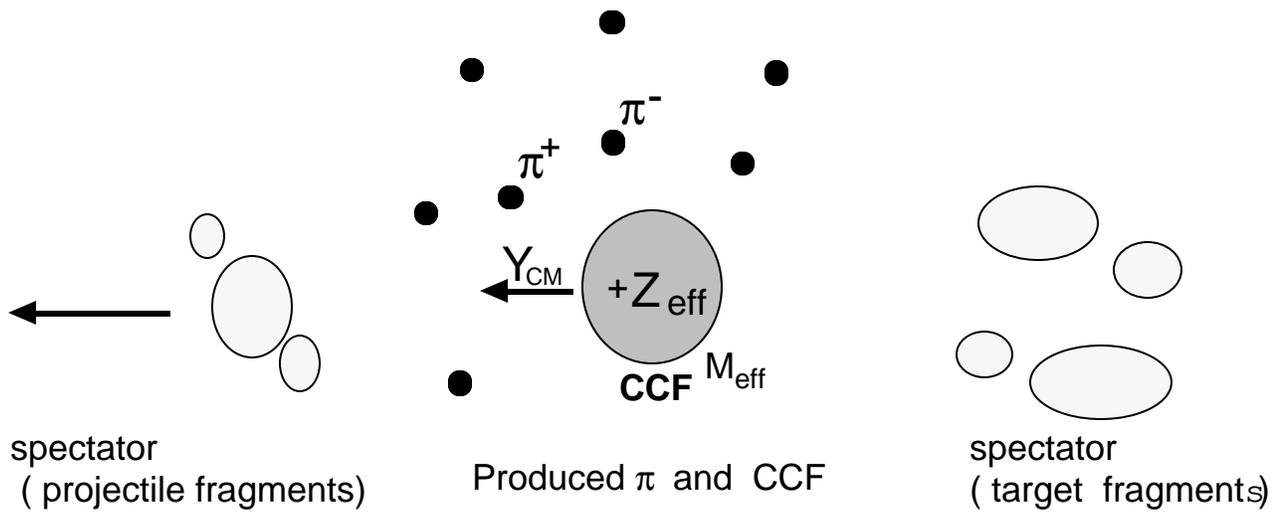
\item[{\bf Fig. 1.}] The picture of collision: charged $\pi$'s and
CCF are produced in the central rapidity region whereas spectators
populate both target and projectile fragmentation regions. The
transverse energy spectra of the charged $\pi$'s are affected by
their Coulomb interaction with the CCF. 
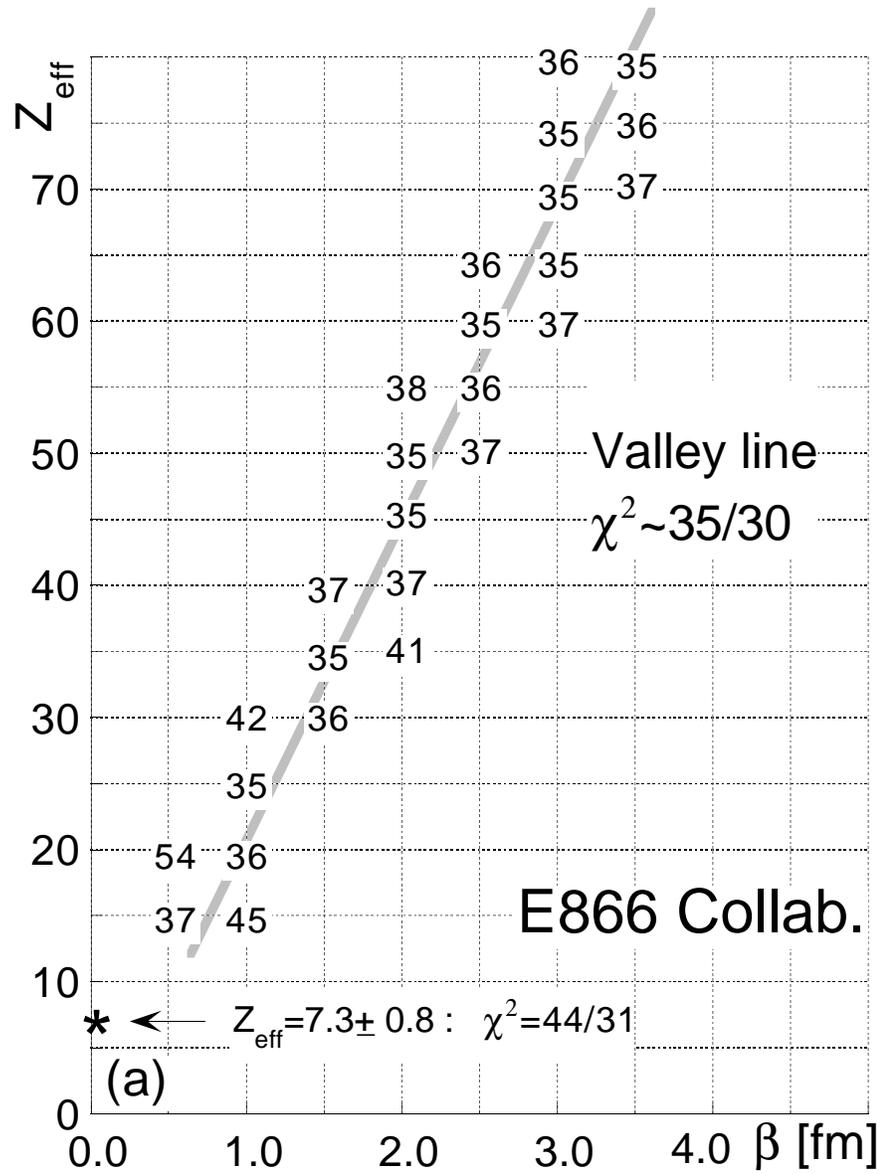
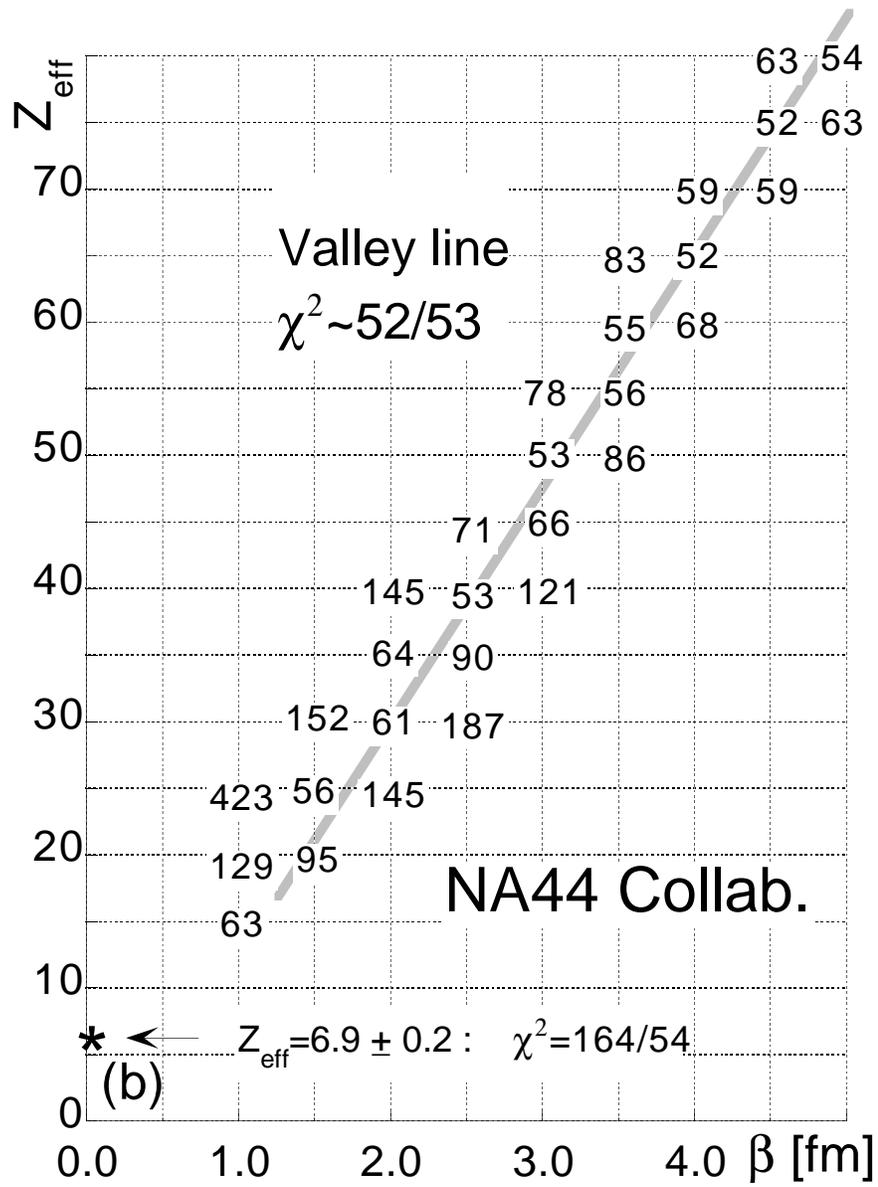
\item[{\bf Fig. 2.}] $(a)$ The valley-like structure exhibited by
values of $\chi^2$ (only numerators are shown, 
denominators, denoting the NDF's, are equal to $30$ here) 
in the $\beta$-$Z_{\mbox{\scriptsize eff}}$ parameters space
for E866 data \cite{E866AuAu} fitted by using eq.(\ref{FINAL}). The
minimum $\chi^2$-value is about 35/30. $(b)$ The same as for $(a)$
but for NA44 data \cite{NA44PbPb} (here the NDF's are equal to
$53$). The minimum $\chi^2$-value is about 52/53. In both
cases the $Z_{\mbox{\scriptsize eff}}$ from fits using Gamow factor
only are shown by the star ($Z_{\mbox{\scriptsize eff}}=7.3$ for
$\chi^2=44/31$ for $(a)$ and $Z_{\mbox{\scriptsize eff}}=6.9$ for
$\chi^2=164/54$ for $(b)$). For E866 data eq.(\ref{FINAL}) has been
renormalized accordingly \cite{F2}.
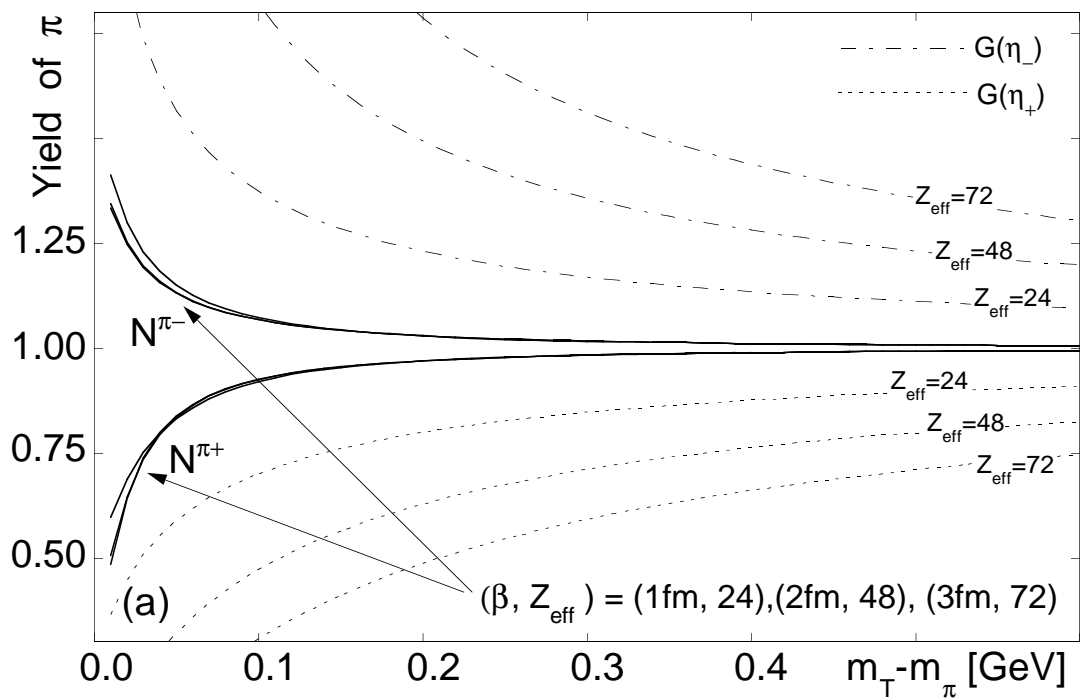
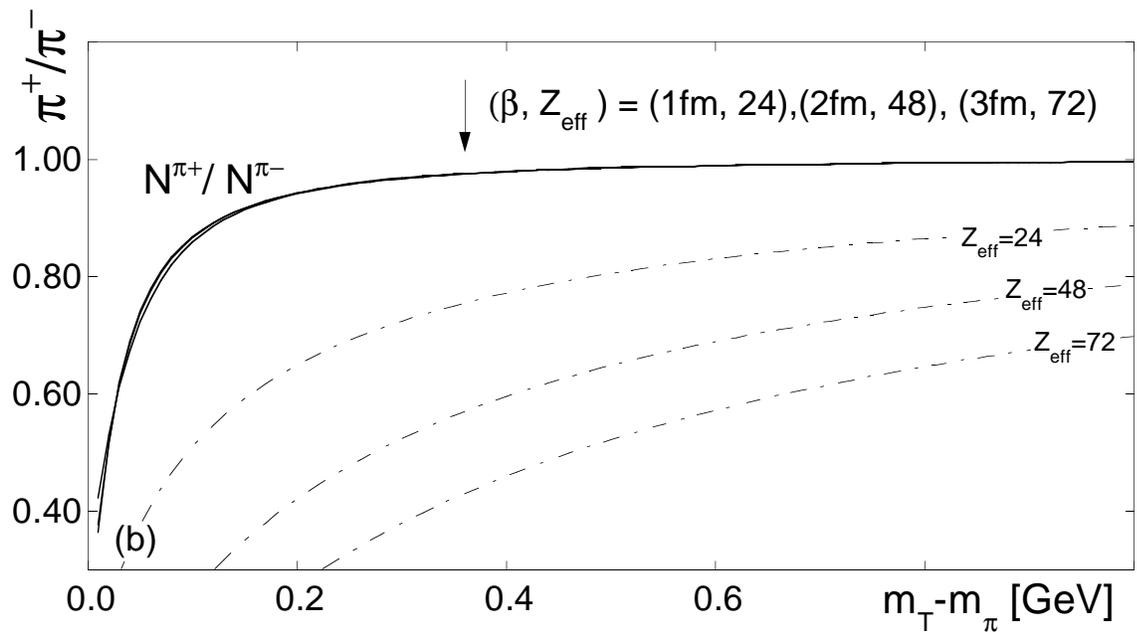
\item[{\bf Fig. 3.}] The quasi-scaling property of eq.(\ref{WAVE-INTEGRAL})
for: $(a)$ $N^{\pi^-}$ and $N^{\pi^+}$ pion production yields and
$(b)$ for their ratio $N^{\pi^+}/N^{\pi^-}$. Three different (scaled by
factors $\lambda =2$ and $3$) sets of parameters: $(\beta,Z_{\mbox{\scriptsize
eff}})$ =(1.0 fm, 24), (2.0 fm, 48) and (3.0 fm, 72) give
approximately the same results (solid line). Dashed curves represent
the results obtained using Gamow factor. 
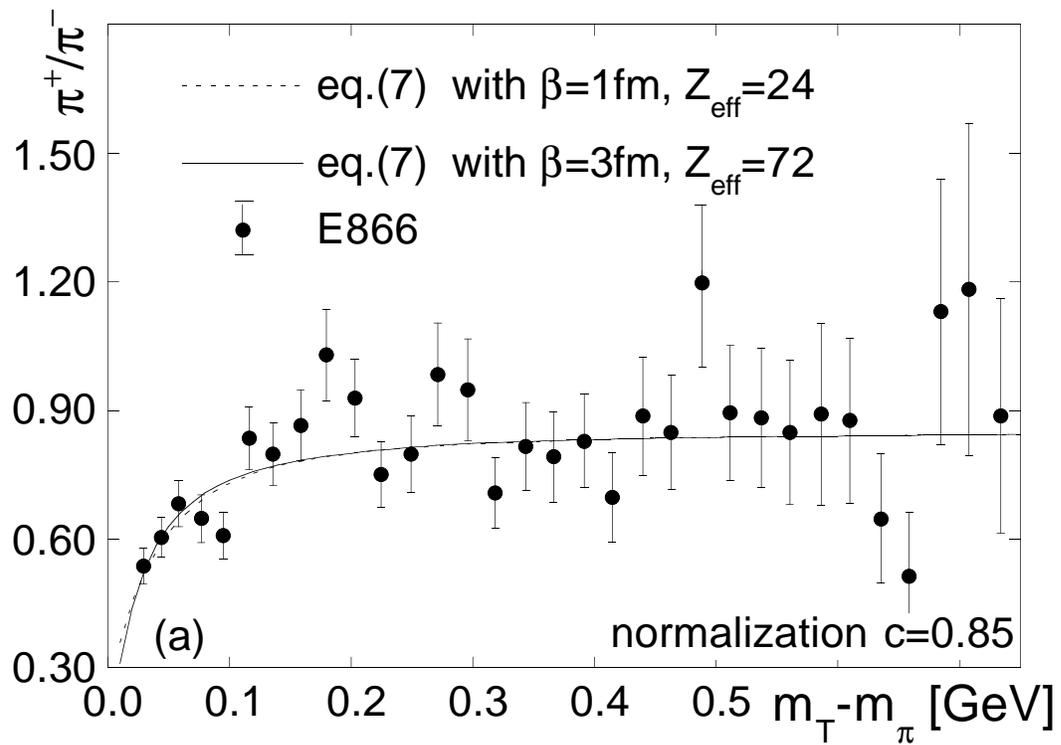
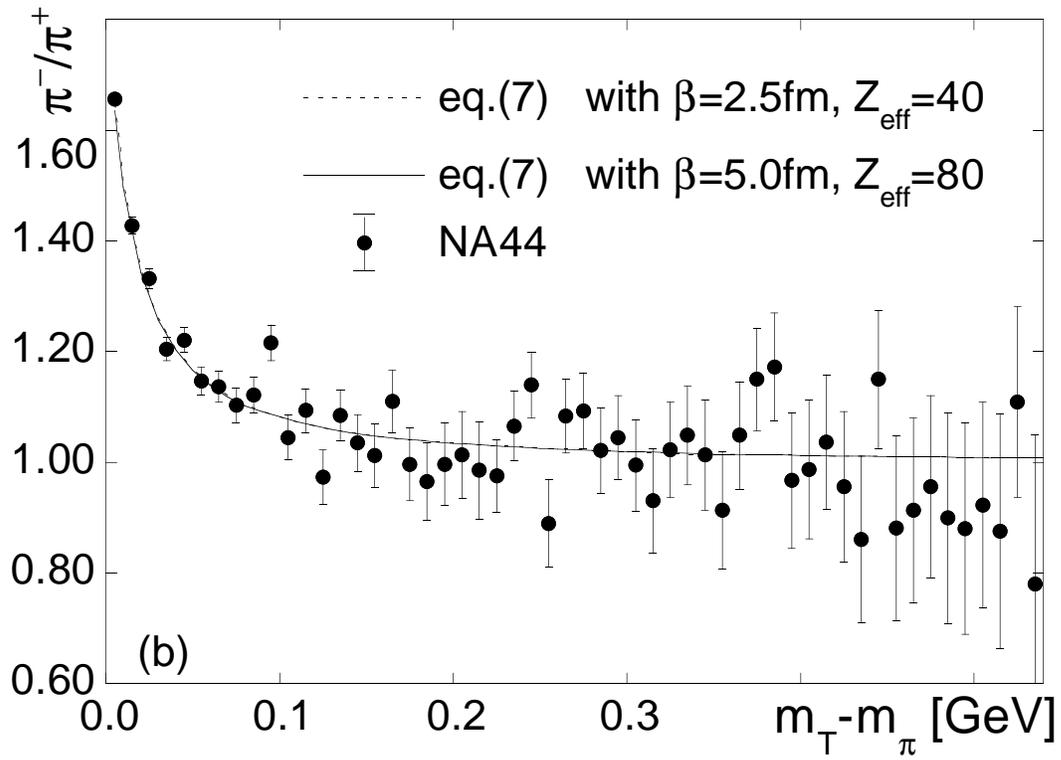
\item[{\bf Fig. 4.}] $(a)$ Comparison of eq.(\ref{FINAL}) with E866
data \cite{E866AuAu} using parameter sets $(\beta,Z_{\mbox{\scriptsize
eff}})$=(1.0 fm, 24) and (3.0 fm, 72). $(b)$ The same for NA44 data
\cite{NA44PbPb} using parameter sets $(\beta,Z_{\mbox{\scriptsize
eff}})$=(2.5 fm, 40) and (5.0 fm, 80). For E866 data eq.(\ref{FINAL})
has been renormalized accordingly \cite{F2}.
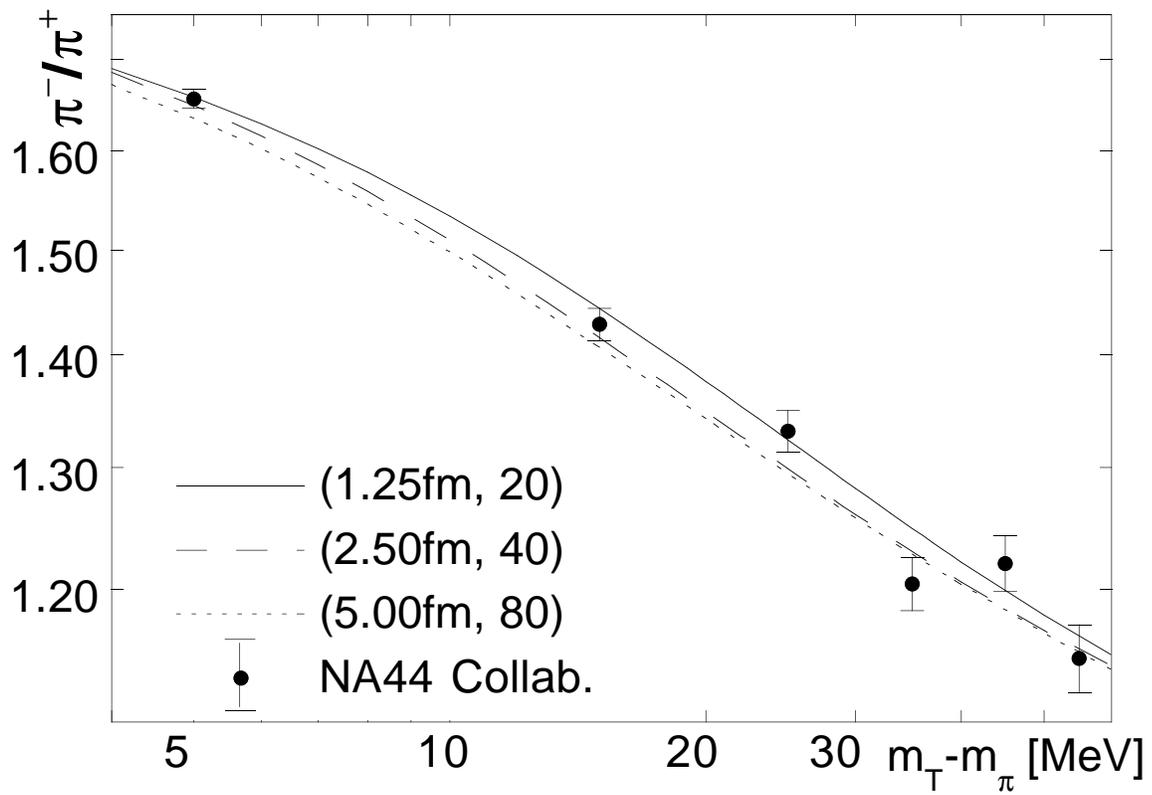
\item[{\bf Fig. 5.}] Example of violation of the quasi-scaling
property of eq.(\ref{WAVE-INTEGRAL}) for small values of the variable
$(m_{\scriptscriptstyle T}-m_{\pi})$ shown for $N^{\pi^+}/N^{\pi^-}$
and compared to NA44 data \cite{NA44PbPb} (here
$(\beta,Z_{\mbox{\scriptsize eff}})$ = (1.25fm,20), (2.5fm,40) and
(5.0fm,80)). 
\end{description}  

\begin{thebibliography}{99}
\bibitem{F12} In this work we are using, as an example, a Gaussian
source function $\rho(r)$ as given below. We have checked that similar
scaling properties hold also for other source functions discussed in
\cite{Biyaj95-1}.
\bibitem{F1} In the actual calculations we evaluate $\mu$ by using
$M_{\mbox{\scriptsize eff}}=Z_{\mbox{\scriptsize eff}}(m_p+m_n)
\approx 2m_pZ_{\mbox{\scriptsize eff}}$,  where $m_p$ and $m_n$
denote the proton and neutron mass, respectively.
\bibitem{F2} Because asymptotic value of E866 data is equal to 0.85,
in our numerical calculations we have introduced additional
normalization constant equal to 0.85 when analysing these data, i.e.,
eq.(\ref{FINAL}) reads in this case:
\begin{eqnarray*}
 \pi^+/\pi^-=~0.85\times  
\frac{N^{\pi^{+}}(m_{\scriptscriptstyle T}-m_{\pi};~~~\eta_+,\beta~)}
{N^{\pi^{+}}(m_{\scriptscriptstyle T}-m_{\pi};~\eta_-,\beta~)}.
\end{eqnarray*}.
\end{thebibliography}
\end{document}